\documentclass[10pt,twocolumn]{article}

\usepackage[utf8]{inputenc}
\usepackage[T1]{fontenc}
\usepackage{lmodern}
\usepackage[margin=1in]{geometry}
\usepackage{amsmath,amssymb}
\usepackage{graphicx}
\usepackage{booktabs}
\usepackage{multirow}
\usepackage{hyperref}
\usepackage{xcolor}
\usepackage{enumitem}
\usepackage{caption}
\usepackage{subcaption}
\usepackage{array}
\usepackage{tabularx}
\usepackage{longtable}
\usepackage{float}
\usepackage{url}
\usepackage{microtype}
\usepackage[numbers,sort&compress]{natbib}
\usepackage{xspace}
\usepackage{abstract}

\newcommand{\framework}{SecLens-R\xspace}
\newcommand{\seclens}{SecLens\xspace}

\hypersetup{
    colorlinks=true,
    linkcolor=blue!60!black,
    citecolor=blue!60!black,
    urlcolor=blue!60!black
}

\title{\textbf{SecLens: Role-Specific Evaluation of LLMs\\for Security Vulnerability Detection}}

\author{%
\makebox[\textwidth]{%
\begin{tabular}[t]{c}
\textbf{Subho Halder\thanks{Corresponding author. Email: \texttt{subho.halder@gmail.com}}} \\
Mattersec Labs
\end{tabular}\hfill
\begin{tabular}[t]{c}
\textbf{Siddharth Saxena} \\
Mattersec Labs
\end{tabular}\hfill
\begin{tabular}[t]{c}
\textbf{Kashinath Kadaba Shrish} \\
Kalmantic Labs
\end{tabular}\hfill
\begin{tabular}[t]{c}
\textbf{Thiyagarajan M} \\
Kalmantic Labs
\end{tabular}%
}%
}

\date{}

\begin{document}

\twocolumn[
  \maketitle
  \begin{onecolabstract}
Existing benchmarks for LLM-based vulnerability detection reduce a model's capability to one number. That number cannot serve the divergent needs of a CISO who prioritizes critical-vulnerability recall, an engineering leader who optimizes for low false-positive rates, or an AI officer who weighs cost against capability. We introduce \framework, a multi-stakeholder evaluation framework built on 35 shared dimensions across 7 measurement categories. Five role-specific weight profiles (CISO, Chief AI Officer, Security Researcher, Head of Engineering, and AI-as-Actor) each select 12--16 dimensions with weights summing to 80, producing a composite Decision Score between 0 and 100. We evaluate 12 frontier models on a dataset of 406 tasks drawn from 93 open-source projects spanning 10 programming languages and 8 OWASP-aligned vulnerability categories, using both Code-in-Prompt (CIP) and Tool-Use (TU) evaluation layers. Decision Scores diverge by up to 31 points across roles for the same model: Qwen3-Coder earns an A (76.3) for Head of Engineering but a D (45.2) for CISO; GPT-5.4 earns an A (76.7) for Head of Engineering but a D (48.4) for CISO. These results confirm that model selection for security vulnerability detection is not a single-objective problem, and that stakeholder-aware evaluation surfaces information that aggregate scores cannot.
  \end{onecolabstract}
  \vspace{1.5em}
]
\saythanks

% ============================================================
% 1. INTRODUCTION
% ============================================================
\section{Introduction}
\label{sec:introduction}

Large language models are increasingly applied to security vulnerability detection~\citep{pearce2022asleep,tony2023llmseceval,fang2024llmagents,sheng2025llmsurvey}. Models such as GPT-5.4~\citep{openai2025gpt54}, Claude Sonnet 4.6 and Opus 4.6~\citep{anthropic2025claude46}, and Gemini 3.x~\citep{google2025gemini3} can identify, classify, and localize vulnerabilities across diverse programming languages and CWE categories. The pace of benchmark development has accelerated in parallel: from CyberSecEval~\citep{bhatt2023cyberseceval} and PrimeVul~\citep{ding2024primevul} in 2023--2024, to SEC-bench~\citep{lee2025secbench}, SecVulEval~\citep{lu2025secvuleval}, and TOSSS~\citep{damie2026tosss} in 2025--2026. Organizations now face a practical question: which model should they deploy?

These benchmarks share a critical limitation: they collapse a model's rich performance profile into a single aggregate score. While such scores are useful for leaderboard rankings, they are insufficient for the decisions organizations actually face.

Consider the divergent needs of different organizational stakeholders:

\begin{itemize}[noitemsep,topsep=2pt]
    \item A \textbf{Chief Information Security Officer (CISO)} asks: ``Can I trust this model in my security program?'' The CISO prioritizes low false-negative rates on critical vulnerabilities, severity-weighted recall, and consistency across CWE categories. A model that excels at injection detection but silently misses authentication bypasses is unacceptable.
    \item A \textbf{Chief AI Officer (CAIO)} asks: ``Which model unlocks new capabilities while balancing risk and cost?'' The CAIO cares about MCC-per-dollar, autonomous completion rates, and tool-use effectiveness at scale.
    \item A \textbf{Security Researcher} asks: ``How deep and reliable is this model's vulnerability reasoning?'' Researchers need CWE taxonomy mastery, evidence chain completeness, and reasoning quality on both true positives and false positives.
    \item A \textbf{Head of Engineering} asks: ``Will this help or hurt my team's velocity and code quality?'' Engineering leaders optimize for high precision (low false-positive rates), fast wall times, low cost per task, and actionable findings with CWE and location.
    \item An \textbf{AI-as-Actor} evaluation asks: ``Does the agent know what it can and can't do?'' This lens evaluates parse reliability, format compliance, error handling, autonomous completion, and graceful degradation.
\end{itemize}

A single benchmark score cannot serve all five perspectives. A model ranked first overall might be the worst choice for a CISO if it misses critical vulnerabilities, or the worst choice for engineering if it generates excessive false positives. Our empirical results confirm this: the same model can score 31 points apart depending on which stakeholder lens is applied.

\subsection{Contributions}

We make the following contributions:

\begin{enumerate}[noitemsep,topsep=2pt]
    \item \textbf{35 shared evaluation dimensions} across 7 categories, applied through 5 role-specific weight profiles where each role selects 12--16 dimensions with weights summing to 80.
    \item \textbf{A weighted composite scoring methodology} with dynamic dimension exclusion: when a dimension is unavailable (e.g., tool-use metrics for Code-in-Prompt runs), the denominator adjusts automatically.
    \item \textbf{Four normalization strategies} with fixed reference caps, eliminating cohort-relative scoring artifacts.
    \item \textbf{Integration with \seclens}, a two-layer benchmark evaluating LLMs on vulnerability detection using both Code-in-Prompt (CIP) and Tool-Use (TU) paradigms.
    \item \textbf{Empirical validation across 12 frontier models} on 406 tasks, demonstrating Decision Score divergences of up to 31 points and confirming that leaderboard rank does not predict role-specific rank.
\end{enumerate}

The remainder of this paper is organized as follows. Section~\ref{sec:related} reviews related work. Section~\ref{sec:framework} presents the framework design. Section~\ref{sec:catalog} provides the dimension catalog. Section~\ref{sec:methodology} describes the evaluation methodology. Section~\ref{sec:experiments} details the experimental design. Section~\ref{sec:analysis} presents results and analysis. Section~\ref{sec:discussion} addresses limitations and future work. Section~\ref{sec:conclusion} concludes.

% ============================================================
% 2. RELATED WORK
% ============================================================
\section{Related Work}
\label{sec:related}

\subsection{LLM Security Benchmarks}

Several benchmarks evaluate LLM capabilities in security contexts. CyberSecEval~\citep{bhatt2023cyberseceval} from Meta evaluates both the tendency of LLMs to generate insecure code and their ability to assist in cyberattacks. CyberSecEval~2~\citep{bhatt2024purplellama} adds prompt injection resistance and code interpreter abuse. CyberSecEval~3~\citep{wan2024cyberseceval3} extends to offensive capability evaluation. SecEval~\citep{li2023seceval} covers code generation security across 130 CWE categories. SWE-bench~\citep{jimenez2024swebench} evaluates models on real-world GitHub issues, though it targets bug-fixing rather than vulnerability detection.

More recent work has sharpened the picture. PrimeVul~\citep{ding2024primevul} demonstrates that existing benchmarks dramatically overestimate vulnerability detection performance (68\% F1 on BigVul drops to 3\% on their deduplicated dataset). VulBench~\citep{gao2023vulbench} aggregates CTF and real-world CVEs with root cause annotations. VulDetectBench~\citep{liu2024vuldetectbench} evaluates 17 models across tasks of increasing difficulty, finding that models achieve 80\%+ on binary identification but under 30\% on detailed analysis. SecLLMHolmes~\citep{ullah2024secllmholmes} shows that LLM vulnerability reasoning is non-deterministic and sensitive to variable naming. IRIS~\citep{li2024iris} combines LLMs with static analysis to detect 103\% more vulnerabilities than CodeQL alone. SASTBench~\citep{feiglin2025sastbench} evaluates LLM agents on SAST false-positive triage using real CVEs. A recent survey by \citet{sheng2025llmsurvey} provides a structured overview of LLM architectures, fine-tuning strategies, and evaluation metrics for vulnerability detection.

Three concurrent benchmarks are particularly relevant. SecVulEval~\citep{lu2025secvuleval} is the largest CVE-grounded benchmark to date, covering 25,440 C/C++ function samples across 5,867 CVEs with statement-level ground truth; the best model (Claude~3.7 Sonnet) achieves only 23.83\% F1, underscoring the difficulty of the task. SEC-bench~\citep{lee2025secbench} introduces automated CVE reproduction with proof-of-concept generation and patch validation, evaluating LLM agents on 200 verified instances; top models achieve at most 18\% PoC generation and 34\% patching success. TOSSS~\citep{damie2026tosss}, published concurrently with our work, frames vulnerability detection as binary snippet selection across C/C++ and Java CVEs; model scores range from 0.48 to 0.89.

LLMSecEval~\citep{tony2023llmseceval} provides security-relevant code completions based on MITRE CWE scenarios. SecurityEval~\citep{siddiq2022securityeval} offers a focused dataset for evaluating LLM-generated code.

All of these benchmarks produce single aggregate scores or per-category breakdowns without stakeholder-specific interpretation. None addresses the organizational decision context in which evaluation results are consumed. Our work is complementary: the \framework scoring layer can consume output from any of these benchmarks, transforming a single leaderboard into five role-specific evaluations.

\subsection{Role-Based Evaluation in Software Engineering}

The notion that different stakeholders require different evaluation lenses is well-established in software engineering. ISO/IEC 25010~\citep{iso25010} defines software quality from multiple viewpoints including users, developers, and operators. The Goal-Question-Metric (GQM) paradigm~\citep{basili1994gqm} formalizes the idea that metrics should be derived from stakeholder goals. In testing, risk-based approaches~\citep{amland1999riskbased} weight test cases by organizational impact rather than code coverage alone.

In the machine learning evaluation literature, Model Cards~\citep{mitchell2019modelcards} and Datasheets for Datasets~\citep{gebru2021datasheets} advocate for multi-stakeholder transparency, but focus on documentation rather than providing different quantitative scores for different roles. HELM~\citep{liang2022helm} evaluates LLMs across multiple scenarios and metrics but does not aggregate by stakeholder role. Chatbot Arena~\citep{chiang2024arena} uses crowdsourced preferences to produce Elo ratings, a single-dimensional ranking that our work explicitly moves beyond. AgentBench~\citep{liu2024agentbench} evaluates LLMs as agents across diverse tasks, relevant to our AI-as-Actor lens but without role-specific scoring.

\subsection{Multi-Stakeholder Decision Frameworks}

Multi-criteria decision analysis (MCDA)~\citep{velasquez2013mcda} provides formal methods for aggregating multiple evaluation dimensions with role-specific weight vectors. The Analytic Hierarchy Process (AHP)~\citep{saaty1990ahp} and TOPSIS~\citep{hwang1981topsis} are widely used MCDA techniques for technology selection decisions. Recent work on cost-aware LLM evaluation~\citep{sun2024cebench} demonstrates the practical need for frameworks that consider inference cost alongside quality. Our approach draws on these methods while adapting them to LLM security evaluation: a shared pool of 35 dimensions (rather than per-role dimension sets) simplifies maintenance and enables direct comparison, while role-specific weight vectors preserve stakeholder-specific priorities.

% ============================================================
% 3. FRAMEWORK DESIGN
% ============================================================
\section{Framework Design}
\label{sec:framework}

\subsection{Stakeholder Roles}
\label{sec:roles}

We define five stakeholder roles, each representing a distinct decision context. Table~\ref{tab:roles} summarizes the roles and their core decision questions.

\begin{table}[t]
\centering
\caption{Stakeholder roles and their core evaluation questions.}
\label{tab:roles}
\small
\begin{tabular}{@{}p{2.0cm}p{5.0cm}@{}}
\toprule
\textbf{Role} & \textbf{Core Decision Question} \\
\midrule
CISO & ``Can I trust this model in my security program?'' \\
\addlinespace[3pt]
CAIO / Head of AI & ``Which model unlocks new capabilities while balancing risk and cost?'' \\
\addlinespace[3pt]
Security Researcher & ``Does this model genuinely understand vulnerability mechanics?'' \\
\addlinespace[3pt]
Head of Engineering & ``Will this help or hurt my team's velocity and code quality?'' \\
\addlinespace[3pt]
AI as Actor & ``Does the agent know what it can and can't do?'' \\
\bottomrule
\end{tabular}
\end{table}

\textbf{CISO.} The Chief Information Security Officer is responsible for the organization's security posture, regulatory compliance, and risk management. This lens selects 16 dimensions and allocates 34 of its 80 weight points to Detection (D1/10, D2/8, D3/6, D6/5, D8/5), 18 to Risk \& Severity (D28/10, D29/8), and smaller allocations to Coverage, Reasoning, Efficiency, and Robustness. The CISO accepts higher costs and lower throughput in exchange for trustworthy, severity-aware coverage.

\textbf{CAIO / Head of AI.} The Chief AI Officer evaluates models from a strategic capability and efficiency perspective. This lens selects 14 dimensions, distributing weight across Robustness (D31/4, D32/6, D34/10 = 20), Efficiency (D18/5, D20/8, D22/6 = 19), Detection (D1/9, D4/7 = 16), and Tool-Use (D25/5, D26/3, D27/7 = 15). The CAIO values tool-use effectiveness and autonomous completion as indicators of deployment readiness.

\textbf{Security Researcher.} The security researcher requires deep, technically rigorous evaluation. This lens selects 13 dimensions and concentrates 39 weight points on Detection (D1/8, D2/6, D6/12, D7/10, D8/3) and 21 on Reasoning (D14/10, D15/2, D16/7, D17/2). CWE accuracy (D6, weight 12) is the single heaviest dimension in any profile, reflecting the researcher's need for precise vulnerability classification.

\textbf{Head of Engineering.} The engineering leader optimizes for developer experience and CI/CD integration. This lens selects 13 dimensions and emphasizes Detection (D2/5, D3/12, D7/8, D8/10 = 35) and Efficiency (D18/7, D21/7, D22/5, D23/3 = 22). Precision (D3, weight 12) and Actionable Finding Rate (D8, weight 10) are the top-weighted dimensions because false positives erode developer trust and findings without location data are not actionable.

\textbf{AI as Actor.} This lens evaluates the model's fitness for autonomous operation. It selects 13 dimensions and places 36 of 80 weight points on Robustness (D31/3, D32/6, D33/6, D34/12, D35/9), 18 on Tool-Use (D25/5, D26/5, D27/8), and 17 on Detection (D1/10, D4/7). Autonomous Completion (D34, weight 12) and Graceful Degradation (D35, weight 9) are the distinctive dimensions: the agent must operate without crashing and maintain performance on both common and rare vulnerability classes.

\begin{figure}[t]
\centering
\includegraphics[width=\linewidth]{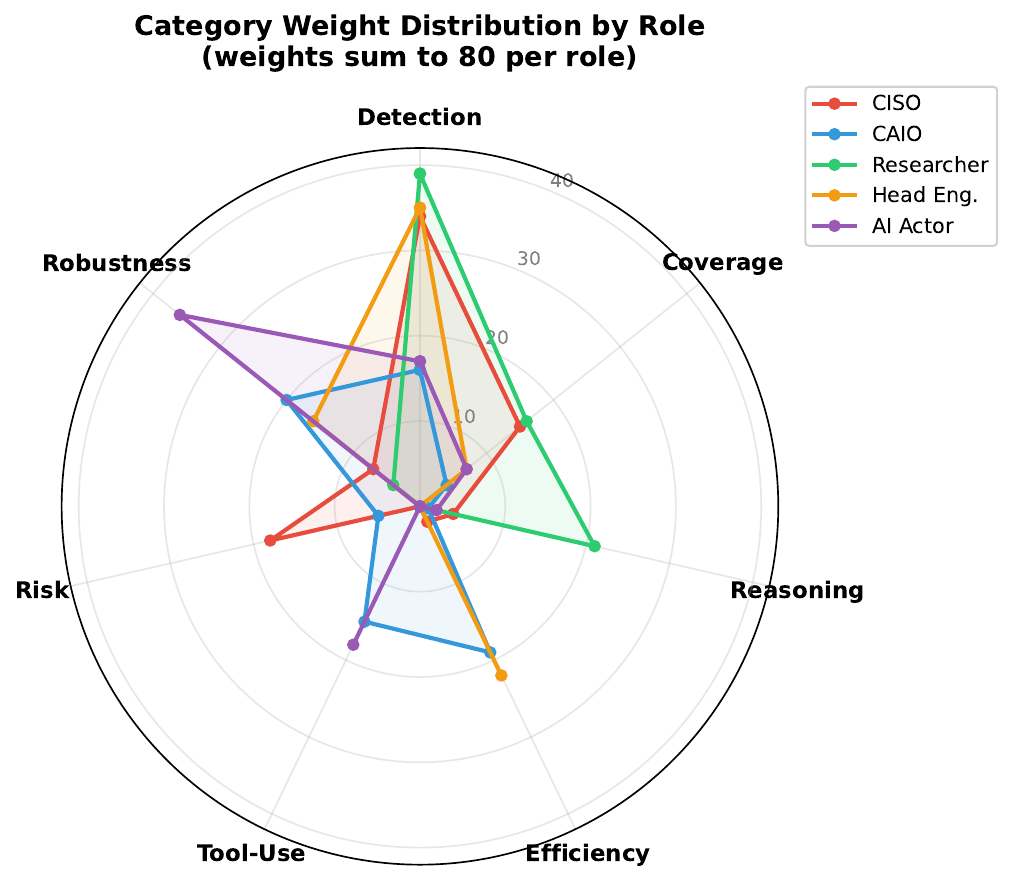}
\caption{Category-level weight distribution for the five stakeholder roles. Each axis represents one of the 7 measurement categories (weights sum to 80 per role).}
\label{fig:radar}
\end{figure}

Figure~\ref{fig:radar} visualizes the category-level weight distribution for each role. The five profiles form distinct shapes: the CISO and Researcher concentrate on Detection and Reasoning, the Head of Engineering balances Detection with Efficiency and Robustness, the CAIO distributes weight broadly, and the AI Actor concentrates heavily on Robustness.

\subsection{Evaluation Layers}
\label{sec:layers}

\framework builds on the \seclens benchmark~\citep{halder2026seclens}, which evaluates models in two layers:

\begin{itemize}[noitemsep,topsep=2pt]
    \item \textbf{Layer 1, Code-in-Prompt (CIP):} The vulnerable function is provided directly in the prompt. The model must identify whether the code is vulnerable, classify the CWE, and provide analysis from a single code snippet. This tests pure reasoning ability without tool use.
    \item \textbf{Layer 2, Tool-Use (TU):} The model is given access to a sandboxed repository clone and three tools: \texttt{read\_file}, \texttt{search}, and \texttt{list\_dir}. This tests real-world auditing ability, requiring multi-turn repository navigation.
\end{itemize}

Each task awards up to 3 points for true positive (vulnerable) tasks: 1 point for correct verdict, 1 for correct CWE identification, and 1 for correct location (file path + line range with IoU above threshold). Negative tasks (post-patch code) award 1 point for correct verdict.

The distinction between CIP and TU is central to the dimension design: several dimensions (D7, D8, D24--D27) are only computable in TU mode, and the scoring formula (Section~\ref{sec:composite}) adjusts automatically when dimensions are unavailable. We use the abbreviations CIP and TU throughout the remainder of this paper.

\subsection{Dimension Taxonomy}
\label{sec:taxonomy}

All five roles draw from a shared pool of 35 dimensions organized into 7 categories. Table~\ref{tab:categories} presents the category taxonomy.

\begin{table}[t]
\centering
\caption{Seven measurement categories and their constituent dimensions.}
\label{tab:categories}
\small
\begin{tabular}{@{}clp{3.0cm}@{}}
\toprule
\textbf{ID} & \textbf{Category} & \textbf{Dimensions} \\
\midrule
A & Detection & D1--D8 \\
B & Coverage \& Consistency & D9--D13 \\
C & Reasoning \& Evidence & D14--D17 \\
D & Operational Efficiency & D18--D23 \\
E & Tool-Use \& Navigation & D24--D27 \\
F & Risk \& Severity & D28--D30 \\
G & Robustness & D31--D35 \\
\bottomrule
\end{tabular}
\end{table}

The shared-pool design means that when two roles both include a dimension (e.g., D1: MCC), they measure the same underlying quantity but assign different weights. This enables direct comparison of how weight allocation, rather than measurement definition, drives divergent evaluations.

\subsection{Weight Assignment Methodology}
\label{sec:weights}

Each role has a weight vector $\mathbf{w}^{(r)} = (w_1^{(r)}, w_2^{(r)}, \ldots)$ over a selected subset $S^{(r)} \subseteq \{D1, \ldots, D35\}$ with $|S^{(r)}| \in [12, 16]$. Weights satisfy:

\begin{equation}
\sum_{i \in S^{(r)}} w_i^{(r)} = 80 \quad \forall \, r
\label{eq:weight_constraint}
\end{equation}

Weight assignment followed a structured process:

\begin{enumerate}[noitemsep,topsep=2pt]
    \item \textbf{Role profiling.} For each role, we identified the top decision criteria based on published job descriptions, industry frameworks (NIST CSF, ISO 27001, OWASP), and practitioner judgment.
    \item \textbf{Dimension selection.} Each role selects 12--16 dimensions from the shared pool of 35.
    \item \textbf{Weight distribution.} The 80 weight points are distributed among the selected dimensions, with higher weights for dimensions central to the role's decision question.
    \item \textbf{Verification.} The final vector is verified to sum to exactly 80.
\end{enumerate}

Table~\ref{tab:category_weights} presents the category-level weight distribution. The profiles are defined in YAML files, making them straightforward to customize for organizational needs.

\begin{table*}[t]
\centering
\caption{Category-level weight distribution across roles (weights sum to 80 per role).}
\label{tab:category_weights}
\small
\begin{tabular}{@{}l r r r r r@{}}
\toprule
\textbf{Category} & \textbf{CISO} & \textbf{CAIO} & \textbf{Researcher} & \textbf{Head Eng.} & \textbf{AI Actor} \\
\midrule
A: Detection                 & 34 & 16 & 39 & 35 & 17 \\
B: Coverage \& Consistency   & 15 &  4 & 16 &  7 &  7 \\
C: Reasoning \& Evidence     &  4 &  1 & 21 &  0 &  2 \\
D: Operational Efficiency    &  2 & 19 &  0 & 22 &  0 \\
E: Tool-Use \& Navigation    &  0 & 15 &  0 &  0 & 18 \\
F: Risk \& Severity          & 18 &  5 &  0 &  0 &  0 \\
G: Robustness                &  7 & 20 &  4 & 16 & 36 \\
\midrule
\textbf{Total}               & \textbf{80} & \textbf{80} & \textbf{80} & \textbf{80} & \textbf{80} \\
\bottomrule
\end{tabular}
\end{table*}

Several patterns emerge. The CISO allocates the most weight to Detection (34) and Risk \& Severity (18), reflecting the primacy of trustworthy, severity-aware threat detection. The CAIO balances Robustness (20), Efficiency (19), Detection (16), and Tool-Use (15), seeking models that are both capable and cost-effective. The Security Researcher concentrates on Detection (39) and Reasoning (21), valuing deep vulnerability understanding. The Head of Engineering distributes weight between Detection (35) and Efficiency (22), prioritizing actionable, fast, affordable scanning. The AI-as-Actor lens heavily weights Robustness (36) and Tool-Use (18), testing whether the model can operate autonomously without failure.

\subsection{Composite Decision Score}
\label{sec:composite}

For a model $m$ evaluated under role $r$, let $s_i^{(m)}$ denote the normalized score (in $[0, 1]$) on dimension $i$, and let $A^{(r)} \subseteq S^{(r)}$ denote the set of \emph{available} dimensions (those for which data exists in the current evaluation). The composite Decision Score is:

\begin{equation}
D^{(r)}(m) = \frac{\displaystyle\sum_{i \in A^{(r)}} w_i^{(r)} \cdot s_i^{(m)}}{\displaystyle\sum_{i \in A^{(r)}} w_i^{(r)}} \times 100
\label{eq:decision_score}
\end{equation}

The denominator uses the sum of available weights rather than the fixed total of 80. This dynamic exclusion handles cases where certain dimensions cannot be computed:

\begin{itemize}[noitemsep,topsep=2pt]
    \item \textbf{CIP layer:} Tool-use dimensions D24--D27 and location dimensions D7, D8 are excluded (no tools, no file-level location in Layer 1).
    \item \textbf{No severity data:} Dimensions D28--D30 are excluded when task severity annotations are absent.
    \item \textbf{No SAST FP tasks:} Dimension D13 is excluded when the dataset contains no SAST false-positive tasks.
\end{itemize}

\textbf{Normalization.} We apply four strategies to map raw dimension values to $[0, 1]$:

\begin{equation}
s_i = \begin{cases}
\text{clamp}(v_i, 0, 1) & \text{if Ratio (D2--D17, D26--D35)} \\[4pt]
\frac{v_i + 1}{2} & \text{if MCC (D1 only)} \\[4pt]
1 - \min\!\left(\frac{v_i}{c_i}, 1\right) & \text{if Lower-is-better (D18,D19,D21,D23--D25)} \\[4pt]
\min\!\left(\frac{v_i}{c_i}, 1\right) & \text{if Higher-is-better (D20,D22)}
\end{cases}
\label{eq:normalization}
\end{equation}

where $v_i$ is the raw value and $c_i$ is a fixed reference cap. The caps are: D18 = \$0.50/task, D19 = \$2.00/TP, D20 = 100 MCC/\$, D21 = 120s, D22 = 60 tasks/min, D23 = 50K tokens, D24 = 30 tool calls, D25 = 20 turns. Fixed caps eliminate cohort-relative normalization artifacts: a model's score does not change when the evaluation cohort changes.

\textbf{Grading Scale.} Decision Scores map to letter grades: A $\geq$ 75, B $\geq$ 60, C $\geq$ 50, D $\geq$ 40, F $<$ 40.

% ============================================================
% 4. DIMENSION CATALOG
% ============================================================
\section{Dimension Catalog}
\label{sec:catalog}

\subsection{Master Dimension Table}

Table~\ref{tab:master_dims} presents all 35 shared dimensions with their category, description, and normalization strategy. Descriptions are drawn from the computational definitions in the implementation.

\begin{table*}[!htbp]
\centering
\caption{Master dimension catalog (35 dimensions). Strategy: R = Ratio, M = MCC, L = Lower-is-better, H = Higher-is-better.}
\label{tab:master_dims}
\footnotesize
\begin{tabular}{@{}r l l p{6.2cm} c@{}}
\toprule
\textbf{ID} & \textbf{Name} & \textbf{Category} & \textbf{Description} & \textbf{Norm.} \\
\midrule
D1  & MCC                        & Detection & Matthews Correlation Coefficient~\citep{matthews1975mcc,chicco2020mcc}; balanced metric robust to class imbalance & M \\
D2  & Recall                     & Detection & True positive rate; fraction of vulnerable code correctly flagged & R \\
D3  & Precision                  & Detection & Positive predictive value; fraction of ``vulnerable'' verdicts that are correct & R \\
D4  & F1                         & Detection & Harmonic mean of precision and recall & R \\
D5  & True Negative Rate         & Detection & Specificity; fraction of non-vulnerable code correctly cleared & R \\
D6  & CWE Accuracy               & Detection & Correct CWE-ID among true positive detections & R \\
D7  & Mean Location IoU          & Detection & Average intersection-over-union of predicted vs.\ ground-truth line ranges (TU only) & R \\
D8  & Actionable Finding Rate    & Detection & Fraction of TPs with verdict + CWE + location (fully actionable) & R \\
\midrule
D9  & CWE Coverage Breadth       & Coverage & Fraction of CWE categories with $\geq$1 correct detection & R \\
D10 & Worst Category Floor       & Coverage & Minimum F1 across all vulnerability categories; no blind spots & R \\
D11 & Cross-Language Consistency  & Coverage & $1 - \text{StdDev}$ of F1 across programming languages & R \\
D12 & Worst Language Floor       & Coverage & Minimum F1 across all programming languages & R \\
D13 & SAST FP Filtering          & Coverage & Accuracy on SAST false-positive tasks (currently excluded; no SAST FP tasks in dataset) & R \\
\midrule
D14 & Evidence Completeness      & Reasoning & Fraction of TPs with source, sink, and data flow evidence & R \\
D15 & Reasoning Presence         & Reasoning & Fraction of all responses with reasoning field populated & R \\
D16 & Reasoning + Correct Verdict & Reasoning & Fraction with reasoning AND correct verdict & R \\
D17 & FP Reasoning Quality       & Reasoning & Among false positive predictions, fraction with reasoning present & R \\
\midrule
D18 & Cost per Task              & Efficiency & Average USD per task & L \\
D19 & Cost per True Positive     & Efficiency & Dollars per correctly detected vulnerability & L \\
D20 & MCC per Dollar             & Efficiency & MCC divided by total cost; quality per dollar & H \\
D21 & Wall Time per Task         & Efficiency & Average seconds per task & L \\
D22 & Throughput                 & Efficiency & Tasks per minute & H \\
D23 & Tokens per Task            & Efficiency & Average total tokens consumed & L \\
\midrule
D24 & Tool Calls per Task        & Tool-Use & Average number of tool invocations (TU only) & L \\
D25 & Turns per Task             & Tool-Use & Average conversation turns (TU only) & L \\
D26 & Navigation Efficiency      & Tool-Use & Fraction of tool calls that accessed relevant files (TU only) & R \\
D27 & Tool Effectiveness         & Tool-Use & Among tool-using tasks, fraction with score $>$ 0 (TU only) & R \\
\midrule
D28 & Severity-Weighted Recall   & Risk & Recall weighted by advisory-reported severity (critical $4\times$, high $3\times$, medium $2\times$, low $1\times$) & R \\
D29 & Critical Miss Rate         & Risk & $1 -$ miss rate on critical/high severity vulnerabilities & R \\
D30 & Severity Coverage          & Risk & Fraction of severity tiers with $\geq$1 correct detection & R \\
\midrule
D31 & Parse Success Rate         & Robustness & Fraction of responses with parseable output (FULL or PARTIAL) & R \\
D32 & Format Compliance          & Robustness & FULL parse rate; fully schema-compliant structured output & R \\
D33 & Error Rate                 & Robustness & $1 -$ fraction of tasks that crashed or produced errors & R \\
D34 & Autonomous Completion      & Robustness & Fraction of tasks completing without error or parse failure & R \\
D35 & Graceful Degradation       & Robustness & $1 - |\text{common\_acc} - \text{rare\_acc}|$; stable performance across common and rare CWEs & R \\
\bottomrule
\end{tabular}
\end{table*}

\subsection{Role Weight Profiles}

Each role selects a subset of dimensions and assigns integer weights summing to 80. Tables~\ref{tab:ciso_profile}--\ref{tab:ai_actor_profile} present the five profiles.

\begin{table}[t]
\centering
\caption{CISO weight profile (16 dimensions, $\Sigma = 80$).}
\label{tab:ciso_profile}
\small
\begin{tabular}{@{}l l r@{}}
\toprule
\textbf{Dim} & \textbf{Name} & \textbf{Wt} \\
\midrule
D1  & MCC                        & 10 \\
D2  & Recall                     & 8 \\
D3  & Precision                  & 6 \\
D5  & True Negative Rate         & 2 \\
D6  & CWE Accuracy               & 5 \\
D8  & Actionable Finding Rate    & 5 \\
D9  & CWE Coverage Breadth       & 4 \\
D10 & Worst Category Floor       & 6 \\
D11 & Cross-Language Consistency  & 3 \\
D14 & Evidence Completeness      & 4 \\
D18 & Cost per Task              & 2 \\
D28 & Severity-Weighted Recall   & 10 \\
D29 & Critical Miss Rate         & 8 \\
D33 & Error Rate                 & 3 \\
D34 & Autonomous Completion      & 3 \\
D35 & Graceful Degradation       & 1 \\
\midrule
    & \textbf{Total}             & \textbf{80} \\
\bottomrule
\end{tabular}
\end{table}

\begin{table}[t]
\centering
\caption{CAIO weight profile (14 dimensions, $\Sigma = 80$).}
\label{tab:caio_profile}
\small
\begin{tabular}{@{}l l r@{}}
\toprule
\textbf{Dim} & \textbf{Name} & \textbf{Wt} \\
\midrule
D1  & MCC                        & 9 \\
D4  & F1                         & 7 \\
D9  & CWE Coverage Breadth       & 4 \\
D15 & Reasoning Presence         & 1 \\
D18 & Cost per Task              & 5 \\
D20 & MCC per Dollar             & 8 \\
D22 & Throughput                 & 6 \\
D25 & Turns per Task             & 5 \\
D26 & Navigation Efficiency      & 3 \\
D27 & Tool Effectiveness         & 7 \\
D30 & Severity Coverage          & 5 \\
D31 & Parse Success Rate         & 4 \\
D32 & Format Compliance          & 6 \\
D34 & Autonomous Completion      & 10 \\
\midrule
    & \textbf{Total}             & \textbf{80} \\
\bottomrule
\end{tabular}
\end{table}

\begin{table}[t]
\centering
\caption{Security Researcher weight profile (13 dimensions, $\Sigma = 80$).}
\label{tab:researcher_profile}
\small
\begin{tabular}{@{}l l r@{}}
\toprule
\textbf{Dim} & \textbf{Name} & \textbf{Wt} \\
\midrule
D1  & MCC                        & 8 \\
D2  & Recall                     & 6 \\
D6  & CWE Accuracy               & 12 \\
D7  & Mean Location IoU          & 10 \\
D8  & Actionable Finding Rate    & 3 \\
D9  & CWE Coverage Breadth       & 7 \\
D10 & Worst Category Floor       & 5 \\
D11 & Cross-Language Consistency  & 4 \\
D14 & Evidence Completeness      & 10 \\
D15 & Reasoning Presence         & 2 \\
D16 & Reasoning + Correct Verdict & 7 \\
D17 & FP Reasoning Quality       & 2 \\
D35 & Graceful Degradation       & 4 \\
\midrule
    & \textbf{Total}             & \textbf{80} \\
\bottomrule
\end{tabular}
\end{table}

\begin{table}[t]
\centering
\caption{Head of Engineering weight profile (13 dimensions, $\Sigma = 80$).}
\label{tab:engineer_profile}
\small
\begin{tabular}{@{}l l r@{}}
\toprule
\textbf{Dim} & \textbf{Name} & \textbf{Wt} \\
\midrule
D2  & Recall                     & 5 \\
D3  & Precision                  & 12 \\
D5  & True Negative Rate         & 4 \\
D7  & Mean Location IoU          & 8 \\
D8  & Actionable Finding Rate    & 10 \\
D12 & Worst Language Floor       & 3 \\
D18 & Cost per Task              & 7 \\
D21 & Wall Time per Task         & 7 \\
D22 & Throughput                 & 5 \\
D23 & Tokens per Task            & 3 \\
D31 & Parse Success Rate         & 7 \\
D32 & Format Compliance          & 3 \\
D33 & Error Rate                 & 6 \\
\midrule
    & \textbf{Total}             & \textbf{80} \\
\bottomrule
\end{tabular}
\end{table}

\begin{table}[t]
\centering
\caption{AI-as-Actor weight profile (13 dimensions, $\Sigma = 80$).}
\label{tab:ai_actor_profile}
\small
\begin{tabular}{@{}l l r@{}}
\toprule
\textbf{Dim} & \textbf{Name} & \textbf{Wt} \\
\midrule
D1  & MCC                        & 10 \\
D4  & F1                         & 7 \\
D9  & CWE Coverage Breadth       & 3 \\
D11 & Cross-Language Consistency  & 4 \\
D14 & Evidence Completeness      & 2 \\
D25 & Turns per Task             & 5 \\
D26 & Navigation Efficiency      & 5 \\
D27 & Tool Effectiveness         & 8 \\
D31 & Parse Success Rate         & 3 \\
D32 & Format Compliance          & 6 \\
D33 & Error Rate                 & 6 \\
D34 & Autonomous Completion      & 12 \\
D35 & Graceful Degradation       & 9 \\
\midrule
    & \textbf{Total}             & \textbf{80} \\
\bottomrule
\end{tabular}
\end{table}

\subsection{Dimension Availability by Layer}

Not all 35 dimensions can be computed in every evaluation setting. In the CIP (Code-in-Prompt) layer, the model receives the vulnerable function directly in the prompt without access to tools or repository navigation. This means:

\begin{itemize}[noitemsep,topsep=2pt]
    \item \textbf{Tool-Use dimensions} (D24--D27) are excluded: no tool calls occur.
    \item \textbf{Location dimensions} (D7, D8) are excluded: CIP provides code inline, so file-level location is not evaluated.
    \item \textbf{SAST FP dimension} (D13) is excluded: the current dataset contains no SAST false-positive tasks.
    \item \textbf{Severity dimensions} (D28--D30) require task-level severity annotations; they are included when severity data is available.
\end{itemize}

When dimensions are excluded, Equation~\ref{eq:decision_score} adjusts the denominator so that the Decision Score reflects only the available evidence. A role that weights excluded dimensions heavily will have a smaller effective denominator, but the score still ranges from 0 to 100.

% ============================================================
% 5. EVALUATION METHODOLOGY
% ============================================================
\section{Evaluation Methodology}
\label{sec:methodology}

\subsection{SecLens Integration}
\label{sec:integration}

As described in Section~\ref{sec:layers}, \framework extends the \seclens benchmark~\citep{halder2026seclens}, which evaluates models across CIP and TU layers with per-task scoring of verdict, CWE, and location. The role-specific dimensions consume the per-task result records produced by \seclens, including:
\begin{itemize}[noitemsep,topsep=2pt]
    \item \textbf{Scores:} verdict (0/1), CWE match (0/1), location match (0/1), total earned
    \item \textbf{Parse result:} status (FULL/PARTIAL/FAILED), parsed output fields (verdict, CWE, location, reasoning, evidence chain with source/sink/flow)
    \item \textbf{Metrics:} cost\_usd, total\_tokens, wall\_time\_s, tool\_calls, turns
    \item \textbf{Task metadata:} task\_type, task\_category, task\_language, task\_severity
\end{itemize}

\subsection{Scoring Pipeline}
\label{sec:pipeline}

The end-to-end scoring pipeline proceeds as follows:

\begin{enumerate}[noitemsep,topsep=2pt]
    \item \textbf{Run evaluation.} Execute \seclens with a specified model, dataset, layer, and prompt preset. This produces a JSONL file of per-task \texttt{TaskResult} records.
    \item \textbf{Compute dimensions.} For each of the 35 shared dimensions, compute the raw value from the result records using the dimension functions.
    \item \textbf{Normalize.} Apply the dimension-specific normalization strategy (Ratio, MCC, Lower-is-better, or Higher-is-better) to map each raw value to $[0, 1]$.
    \item \textbf{Select and weight.} For each role, select the role's dimension subset, multiply each normalized score by its role-specific weight, and compute the Decision Score via Equation~\ref{eq:decision_score}.
    \item \textbf{Grade and report.} Map the Decision Score to a letter grade and produce a per-role report with category subtotals and individual dimension scores.
\end{enumerate}

% ============================================================
% 6. EXPERIMENTAL DESIGN
% ============================================================
\section{Experimental Design}
\label{sec:experiments}

\subsection{Models Under Evaluation}
\label{sec:models}

We evaluate 12 models spanning four providers:

\textbf{Anthropic:} Claude Opus 4.6, Claude Sonnet 4.6, Claude Haiku 4.5.

\textbf{Google:} Gemini 3.1 Pro Preview, Gemini 3 Flash Preview, Gemini 2.5 Pro, Gemini 2.5 Flash.

\textbf{OpenAI:} GPT-5.4.

\textbf{Other (via OpenRouter):} Qwen3-Coder, Qwen3-Coder-Plus, Kimi K2.5, Grok Code Fast 1.

This selection spans frontier reasoning models (Opus 4.6, Gemini 3.1 Pro), mid-tier models (Sonnet 4.6, Gemini 2.5 Pro), cost-optimized models (Haiku 4.5, Gemini 2.5 Flash), and open-weight models accessed through third-party routing (Qwen3, Kimi, Grok).

\subsection{Dataset}
\label{sec:dataset}

We use the \seclens dataset, which contains tasks derived from confirmed CVEs across multiple CWE categories and programming languages. Table~\ref{tab:dataset_stats} summarizes the dataset statistics.

\begin{table}[t]
\centering
\caption{Dataset statistics.}
\label{tab:dataset_stats}
\small
\begin{tabular}{@{}l r@{}}
\toprule
\textbf{Attribute} & \textbf{Count} \\
\midrule
Total tasks                    & 406 \\
True Positive tasks            & 203 \\
Post-Patch tasks               & 203 \\
Source projects                & 93 \\
OWASP categories               & 8 \\
Programming languages          & 10 \\
Distinct CVEs                  & $\sim$203 \\
\bottomrule
\end{tabular}
\end{table}

The dataset includes two task types. \textbf{True Positive (TP)} tasks contain pre-patch code with a confirmed vulnerability. \textbf{Post-Patch} tasks contain the same function after the official fix has been applied. Each TP task is paired with its post-patch counterpart, yielding a balanced 203/203 split.

\textbf{Vulnerability categories.} The 8 categories align with the OWASP Top 10:2021 taxonomy~\citep{owasp2021top10}, with 6 direct mappings and 2 extended categories. Table~\ref{tab:categories_data} shows the distribution.

\begin{table}[t]
\centering
\caption{Vulnerability category distribution (TP + Post-Patch).}
\label{tab:categories_data}
\small
\begin{tabular}{@{}p{3.2cm} l r@{}}
\toprule
\textbf{Category} & \textbf{OWASP} & \textbf{Tasks} \\
\midrule
Broken Access Control      & A01:2021 & 82 \\
Cryptographic Failures     & A02:2021 & 64 \\
Injection                  & A03:2021 & 62 \\
Improper Input Validation  & Extended & 58 \\
SSRF                       & A10:2021 & 46 \\
Auth Failures              & A07:2021 & 38 \\
Data Integrity Failures    & A08:2021 & 36 \\
Memory Safety              & Extended & 20 \\
\bottomrule
\end{tabular}
\end{table}

Memory Safety (CWE-119 family, covering C/C++ buffer overflows) and Improper Input Validation (CWE-20 family) extend beyond the OWASP Top 10 to cover vulnerability classes not fully captured by the injection and SSRF categories.

\textbf{Programming languages.} 10 languages are represented: PHP (54 tasks), Go (54), Python (48), C\# (46), Ruby (36), Java (36), C (36), Rust (34), JavaScript (32), and C++ (30).

\textbf{Severity distribution.} Among the 203 TP tasks with advisory-reported severity: Critical (25), High (74), Medium (83), Low (21).

\subsection{Evaluation Protocol}
\label{sec:protocol}

Each model was evaluated under the following conditions:

\begin{itemize}[noitemsep,topsep=2pt]
    \item \textbf{Both layers:} CIP (Code-in-Prompt) and TU (Tool-Use) for all models supporting tool calling.
    \item \textbf{Prompt preset:} \texttt{base} (standard instructions).
    \item \textbf{Mode:} \texttt{guided} (category hint provided).
    \item \textbf{Full dataset:} All 406 tasks per run, seed=42.
\end{itemize}

For each model $\times$ layer combination, we compute all applicable dimensions and the five role-specific Decision Scores. This yields 23 total evaluation runs (not all models completed both layers).

% ============================================================
% 7. RESULTS AND ANALYSIS
% ============================================================
\section{Results and Analysis}
\label{sec:analysis}

\subsection{Overall Leaderboard}
\label{sec:leaderboard}

Table~\ref{tab:leaderboard} presents the full leaderboard across both CIP and TU layers. CIP scores range from 37.3\% (Qwen3-Coder) to 49.6\% (Gemini 3 Flash Preview). TU scores are consistently lower, ranging from 31.1\% (GPT-5.4) to 45.8\% (Gemini 3.1 Pro Preview), reflecting the added difficulty of multi-turn repository navigation.

Cost data was tracked for Anthropic, Google, and OpenAI models. The four OpenRouter-routed models (Kimi K2.5, Grok Code Fast 1, Qwen3-Coder, Qwen3-Coder-Plus) did not have cost tracking available through the provider, so cost-derived dimensions (D18, D19, D20) are excluded from their scoring and marked N/T in cost columns.

\begin{table*}[t]
\centering
\caption{Full leaderboard across CIP and TU layers, sorted by CIP score. Cost shown where tracked; N/T = not tracked.}
\label{tab:leaderboard}
\small
\begin{tabular}{@{}l c r c r@{}}
\toprule
\textbf{Model} & \textbf{CIP Score (\%)} & \textbf{CIP Cost} & \textbf{TU Score (\%)} & \textbf{TU Cost} \\
\midrule
Gemini 3 Flash Preview     & 49.6 & \$15.87   & 44.2 & \$184.21 \\
Gemini 3.1 Pro Preview     & 48.2 & \$44.81   & 45.8 & \$639.09 \\
Claude Sonnet 4.6          & 47.6 & \$6.76    & 42.1 & \$359.76 \\
Kimi K2.5                  & 46.8 & N/T       & --   & -- \\
Gemini 2.5 Pro             & 46.2 & \$13.45   & 35.2 & \$57.91 \\
Grok Code Fast 1           & 44.1 & N/T       & 34.4 & N/T \\
Gemini 2.5 Flash           & 44.3 & \$3.23    & 34.2 & \$4.46 \\
Claude Haiku 4.5           & 43.8 & \$2.11    & 37.8 & \$104.06 \\
Claude Opus 4.6            & 41.7 & \$7.56    & 39.0 & \$371.30 \\
Qwen3-Coder-Plus           & 41.2 & N/T       & 34.9 & N/T \\
GPT-5.4                    & 39.9 & \$2.97    & 31.1 & \$23.16 \\
Qwen3-Coder                & 37.3 & N/T       & 32.5 & N/T \\
\bottomrule
\end{tabular}
\end{table*}

\subsection{Per-Role Decision Scores (CIP Layer)}
\label{sec:per_role}

Table~\ref{tab:decision_scores} presents the Decision Scores across all 12 models and 5 roles for the CIP layer. Each cell shows the letter grade and numeric score. Models are sorted by leaderboard score.

\begin{table*}[t]
\centering
\caption{Per-role Decision Scores (CIP layer). Grade thresholds: A $\geq$ 75, B $\geq$ 60, C $\geq$ 50, D $\geq$ 40, F $<$ 40. Sorted by leaderboard score (LB).}
\label{tab:decision_scores}
\small
\begin{tabular}{@{}l c c c c c c@{}}
\toprule
\textbf{Model} & \textbf{LB \%} & \textbf{CISO} & \textbf{CAIO} & \textbf{Researcher} & \textbf{Head Eng.} & \textbf{AI Actor} \\
\midrule
Gemini 3 Flash Preview   & 49.6 & B (73.3) & B (68.1) & B (71.0) & B (66.2) & A (87.5) \\
Gemini 3.1 Pro Preview   & 48.2 & B (67.5) & B (67.0) & B (65.7) & B (63.8) & A (85.7) \\
Claude Sonnet 4.6        & 47.6 & B (65.7) & B (68.4) & B (64.2) & B (73.9) & A (85.6) \\
Kimi K2.5                & 46.8 & B (68.0) & B (67.8) & B (67.0) & B (65.1) & A (86.4) \\
Gemini 2.5 Pro           & 46.2 & B (66.2) & B (67.9) & B (65.2) & B (71.3) & A (86.3) \\
Grok Code Fast 1         & 44.1 & C (58.7) & B (67.5) & B (60.2) & B (73.0) & A (83.8) \\
Gemini 2.5 Flash         & 44.3 & B (61.3) & B (67.9) & B (61.1) & B (72.3) & A (84.9) \\
Claude Haiku 4.5         & 43.8 & B (71.2) & B (69.1) & B (68.2) & B (73.3) & A (85.9) \\
Claude Opus 4.6          & 41.7 & C (51.0) & B (65.6) & C (55.6) & B (72.9) & A (80.2) \\
Qwen3-Coder-Plus         & 41.2 & C (51.1) & B (68.0) & C (54.2) & A (76.9) & A (81.2) \\
GPT-5.4                  & 39.9 & D (48.4) & B (67.0) & C (54.1) & A (76.7) & A (79.2) \\
Qwen3-Coder              & 37.3 & D (45.2) & B (64.0) & C (52.9) & A (76.3) & A (77.9) \\
\bottomrule
\end{tabular}
\end{table*}

Several patterns emerge from the data:

\textbf{AI Actor is universally lenient.} All 12 models earn an A grade (77.9--87.5). This role places 36 of 80 weight points on Robustness, and most frontier models parse and comply well. The narrow score range (9.6 points from worst to best) indicates that current models have largely solved the robustness challenges this lens measures.

\textbf{CISO is the strictest lens.} Scores range from D (45.2, Qwen3-Coder) to B (73.3, Gemini 3 Flash). This role weights severity-weighted recall (D28, weight 10) and critical miss rate (D29, weight 8), which sharply penalize models that fail to detect high-severity vulnerabilities. Only 8 of 12 models achieve B or higher.

\textbf{Head of Engineering favors different models.} GPT-5.4 (A, 76.7), Qwen3-Coder (A, 76.3), and Qwen3-Coder-Plus (A, 76.9) earn their highest grades under this lens, while scoring C or D for the CISO. This role rewards high precision (D3, weight 12), actionable findings (D8, weight 10), fast wall times (D21, weight 7), and low cost (D18, weight 7). Models with conservative prediction strategies (high precision, lower recall) perform well here.

\textbf{Leaderboard rank does not predict role rank.} Gemini 3 Flash leads both the leaderboard (49.6\%) and the CISO lens (73.3). But Claude Haiku 4.5 ranks only 8th on the leaderboard (43.8\%) yet scores 2nd for the CISO (71.2), a jump of 6 positions. Conversely, GPT-5.4 ranks 11th on the leaderboard but 2nd for Head of Engineering.

\textbf{CAIO scores are remarkably stable.} All 12 models score B (64.0--69.1), the tightest band of any role. The 5.1-point spread suggests that current models have comparable profiles on the balanced mix of efficiency, capability, and robustness dimensions the CAIO weights.

\begin{figure*}[t]
\centering
\includegraphics[width=\textwidth]{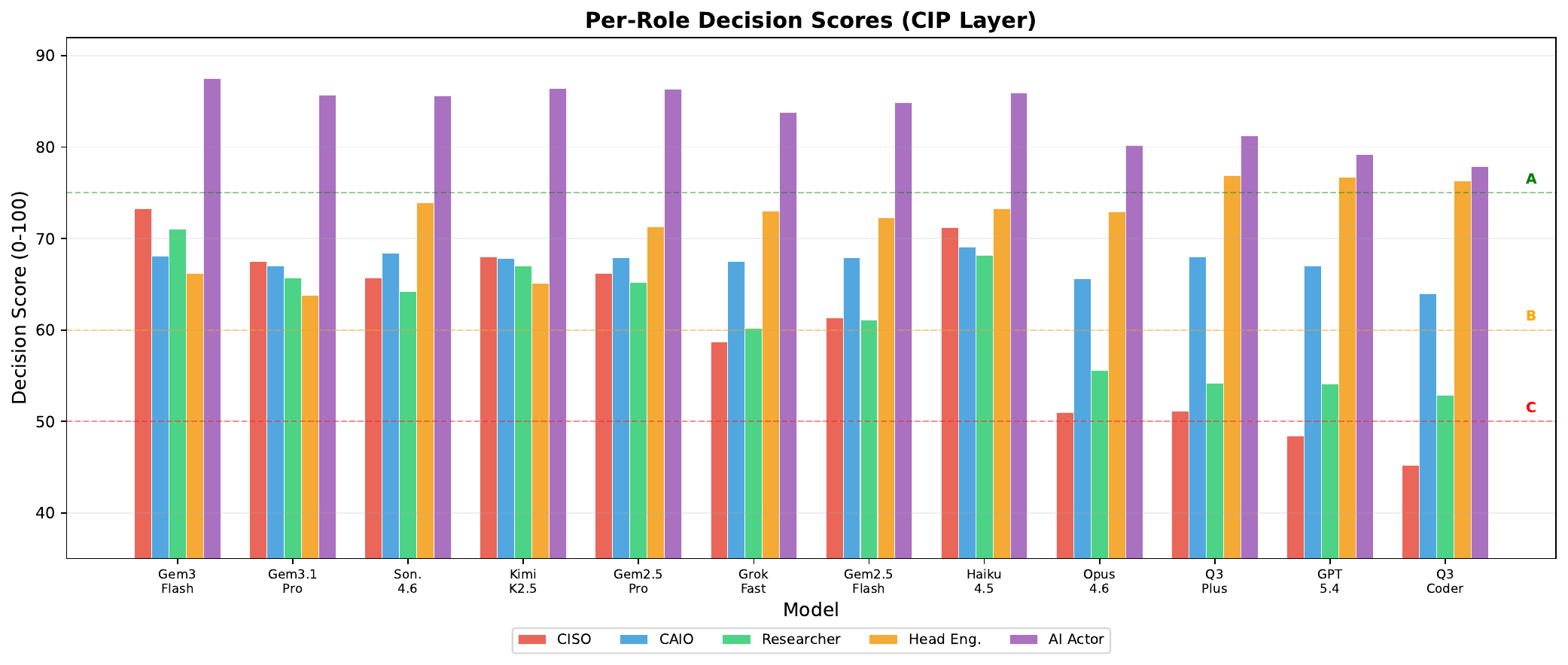}
\caption{Per-role Decision Scores for 12 models (CIP layer). Dashed lines mark grade thresholds: A $\geq$ 75, B $\geq$ 60, C $\geq$ 50. Models sorted by leaderboard score (left to right).}
\label{fig:role_scores}
\end{figure*}

Figure~\ref{fig:role_scores} visualizes these patterns. The AI Actor bars (purple) form a consistently high ceiling, while the CISO bars (red) show the widest variation. The Head of Engineering bars (yellow) rise on the right side of the chart where conservative models (GPT-5.4, Qwen3) appear, crossing over the CISO bars.

\subsection{Per-Category Performance}
\label{sec:per_category}

Table~\ref{tab:category_f1} presents the top-3 and worst models by F1 for each vulnerability category in the CIP layer.

\begin{table*}[t]
\centering
\caption{Per-category F1 leaders and worst performers (CIP layer).}
\label{tab:category_f1}
\small
\begin{tabular}{@{}l l l l l@{}}
\toprule
\textbf{Category} & \textbf{\#1 (F1)} & \textbf{\#2 (F1)} & \textbf{\#3 (F1)} & \textbf{Worst (F1)} \\
\midrule
Broken Access Ctrl.     & Kimi K2.5 (0.667)    & Gemini 3.1 Pro (0.660) & Gemini 3 Flash (0.647) & Qwen3-Coder (0.128) \\
Cryptographic Failures  & Gemini 3 Flash (0.676) & Gemini 3.1 Pro (0.667) & Haiku 4.5 (0.633) & Qwen3-Coder (0.118) \\
Auth Failures           & Kimi K2.5 (0.585)    & Gemini 2.5 Pro (0.579) & Gemini 3.1 Pro (0.578) & Opus 4.6 (0.000) \\
Input Validation        & Haiku 4.5 (0.675)    & Kimi K2.5 (0.649)      & Gemini 2.5 Flash (0.638) & Qwen3-Coder (0.125) \\
Injection               & Gemini 3.1 Pro (0.632) & Gemini 3 Flash (0.618) & Gemini 2.5 Flash (0.603) & Qwen3-Coder (0.062) \\
Memory Safety           & Haiku 4.5 (0.690)    & Gemini 3 Flash (0.667) & Gemini 2.5 Flash (0.640) & Qwen3-Coder (0.308) \\
SSRF                    & Sonnet 4.6 (0.690)   & Opus 4.6 (0.689)       & Qwen3-Plus (0.682) & Qwen3-Coder (0.512) \\
Data Integrity          & Gemini 3 Flash (0.680) & Haiku 4.5 (0.679)     & Gemini 2.5 Flash (0.625) & Qwen3-Coder (0.200) \\
\bottomrule
\end{tabular}
\end{table*}

\textbf{No single model dominates all categories.} Six different models lead at least one category. Gemini 3 Flash leads Cryptographic Failures and Data Integrity; Haiku 4.5 leads Memory Safety and Input Validation; Kimi K2.5 leads Broken Access Control and Auth Failures; Gemini 3.1 Pro leads Injection; and Sonnet 4.6 leads SSRF.

\textbf{SSRF is where Claude models excel.} Sonnet 4.6 (0.690) and Opus 4.6 (0.689) rank \#1 and \#2 for Server-Side Request Forgery, while performing worse on other categories. This suggests architecture-specific strengths in reasoning about network request patterns.

\textbf{Authentication Failures is the hardest category.} The best F1 is only 0.585 (Kimi K2.5), and Opus 4.6 scores 0.000 (a complete miss on all authentication-related vulnerabilities in CIP mode). This category likely requires multi-file reasoning about authentication flows that is difficult from a single code snippet.

\textbf{Qwen3-Coder is consistently worst.} It ranks last in 7 of 8 categories. Inspection of its outputs reveals very high precision but near-zero recall: the model rarely predicts ``vulnerable,'' so it misses almost all true positives while avoiding false positives.

\textbf{Haiku 4.5 punches above its weight.} Despite ranking 8th on the overall leaderboard, it leads Memory Safety (0.690) and Input Validation (0.675). The CISO lens captures this: Haiku ranks 2nd (71.2) under the CISO lens because these category strengths align with the CISO's severity-weighted recall priorities.

\textbf{Per-language variation.} Performance also varies by language. For C code, Gemini 3.1 Pro leads (F1 = 0.750) while Qwen3-Coder trails (0.100). For JavaScript, Gemini 3.1 Pro again leads (0.684) while Qwen3-Coder scores 0.000. For Go, Haiku 4.5 leads (0.647) while GPT-5.4 trails (0.278).

Table~\ref{tab:full_category_f1} presents the complete model-by-category F1 matrix, enabling readers to identify which model is strongest for their specific vulnerability categories of concern.

\begin{table*}[!htbp]
\centering
\caption{F1 score by model and vulnerability category (CIP layer). Bold indicates best-in-category. Italics indicates worst. Models sorted by average F1.}
\label{tab:full_category_f1}
\footnotesize
\begin{tabular}{@{}l c c c c c c c c c@{}}
\toprule
\textbf{Model} & \textbf{BAC} & \textbf{Crypto} & \textbf{Auth} & \textbf{Input} & \textbf{Inj.} & \textbf{Mem.} & \textbf{SSRF} & \textbf{D.Int.} & \textbf{Avg} \\
\midrule
Gemini 3 Flash     & 0.647 & \textbf{0.676} & 0.444 & 0.627 & 0.618 & 0.667 & 0.667 & \textbf{0.680} & 0.628 \\
Claude Haiku 4.5   & 0.586 & 0.633 & 0.545 & \textbf{0.675} & 0.507 & \textbf{0.690} & 0.667 & 0.679 & 0.623 \\
Gemini 3.1 Pro     & 0.660 & 0.667 & 0.578 & 0.618 & \textbf{0.632} & 0.583 & 0.600 & 0.500 & 0.605 \\
Kimi K2.5          & \textbf{0.667} & 0.567 & \textbf{0.585} & 0.649 & 0.517 & 0.609 & 0.644 & 0.529 & 0.596 \\
Gemini 2.5 Pro     & 0.635 & 0.562 & 0.579 & 0.627 & 0.585 & 0.526 & 0.643 & 0.514 & 0.584 \\
Claude Sonnet 4.6  & 0.590 & 0.625 & 0.514 & 0.571 & 0.518 & 0.526 & \textbf{0.690} & 0.537 & 0.571 \\
Gemini 2.5 Flash   & 0.457 & 0.500 & 0.191 & 0.638 & 0.603 & 0.640 & 0.623 & 0.625 & 0.535 \\
Grok Code Fast 1   & 0.530 & 0.367 & 0.529 & 0.561 & 0.440 & 0.588 & 0.667 & 0.414 & 0.512 \\
Claude Opus 4.6    & 0.231 & 0.491 & \textit{0.000} & 0.408 & 0.263 & 0.476 & 0.689 & 0.333 & 0.361 \\
Qwen3-Plus         & 0.310 & 0.216 & 0.320 & 0.417 & 0.350 & 0.400 & 0.682 & 0.320 & 0.377 \\
GPT-5.4            & 0.182 & 0.222 & 0.333 & 0.256 & 0.171 & 0.353 & 0.609 & 0.214 & 0.293 \\
Qwen3-Coder        & \textit{0.128} & \textit{0.118} & 0.100 & \textit{0.125} & \textit{0.062} & \textit{0.308} & \textit{0.512} & \textit{0.200} & \textit{0.194} \\
\bottomrule
\end{tabular}
\end{table*}

\begin{figure*}[t]
\centering
\includegraphics[width=0.85\textwidth]{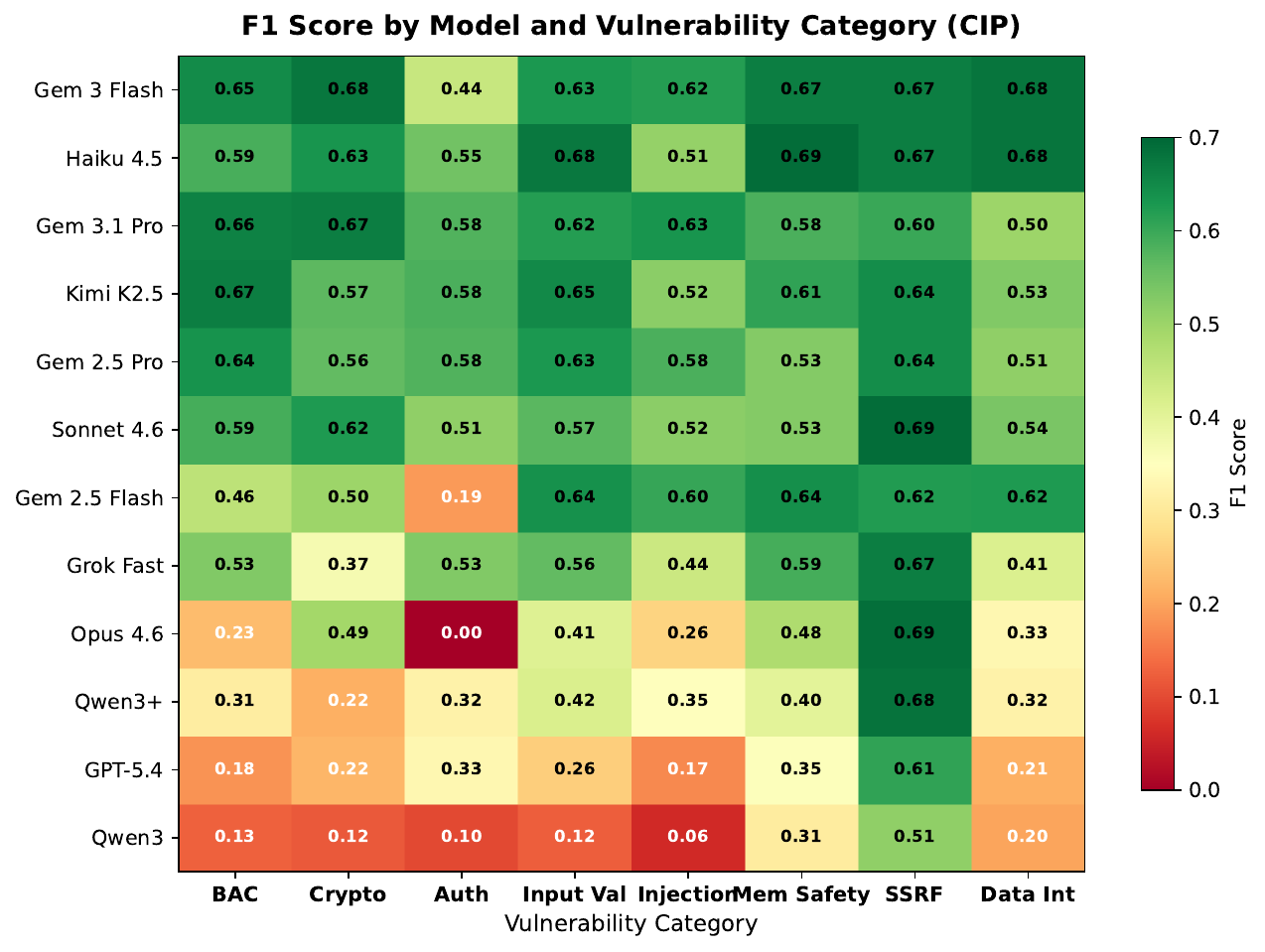}
\caption{F1 score heatmap by model and vulnerability category (CIP layer). Models sorted by average F1. Green = strong, red = weak.}
\label{fig:heatmap}
\end{figure*}

Figure~\ref{fig:heatmap} presents the complete F1 matrix as a heatmap, where green cells indicate strong performance and red cells indicate weakness. The diagonal pattern of strengths across models is visible: no single column is uniformly green, confirming that category-specific evaluation is necessary.

\textbf{Category performance predicts role divergence.} Models with high recall across categories (Gemini 3 Flash, Haiku 4.5) score well for the CISO, whose D28 (Severity-Weighted Recall) and D29 (Critical Miss Rate) aggregate category-level detection. Models with concentrated recall in a few categories but high precision overall (GPT-5.4, Qwen3-Coder) score well for the Head of Engineering, whose D3 (Precision) dominates at weight 12. This table reveals why the same model can score 31 points apart under different lenses: the CISO lens penalizes category-level gaps visible in the Auth Failures and Injection columns, while the Engineering lens rewards the global precision visible in models that rarely flag code as vulnerable.

\subsection{Category-Role Interaction}
\label{sec:cat_role}

The relationship between per-category performance and role scores explains many of the divergences in Table~\ref{tab:decision_scores}. We highlight three instructive cases:

\textbf{Case 1: Claude Haiku 4.5 (CISO 2nd, Leaderboard 8th).} Haiku has the highest recall for Memory Safety (0.909), Input Validation (0.966), and competitive recall across Auth Failures (0.923) and SSRF (0.800). The CISO lens rewards this broad, high-recall coverage through D2 (Recall, weight 8) and D28 (Severity-Weighted Recall, weight 10). Despite lower precision (which the Engineering lens penalizes), Haiku's category-level coverage makes it a strong CISO pick.

\textbf{Case 2: GPT-5.4 (Eng. 2nd, CISO 11th).} GPT-5.4 achieves high precision (0.800 on Broken Access Control, 0.750 on Injection) but very low recall (0.103 and 0.097 respectively). The Engineering lens rewards this through D3 (Precision, weight 12) and D33 (Error Rate, weight 6). The CISO lens penalizes it through D28 (Severity-Weighted Recall), which drops because GPT-5.4 misses the majority of true vulnerabilities including critical ones.

\textbf{Case 3: Opus 4.6 (Auth Failures blind spot).} Opus 4.6 achieves F1 = 0.000 on Authentication Failures (recall 0.000) while performing well on SSRF (F1 = 0.689). The CISO lens captures this via D10 (Worst Category Floor, weight 6), which drives the CISO score down to C (51.0). The AI Actor lens is unaffected because it does not include D10, illustrating how dimension selection creates lens-specific sensitivity to category gaps.

\subsection{Cross-Role Divergence Analysis}
\label{sec:divergence}

We define the \emph{Role Divergence Index} (RDI) for a model $m$:

\begin{equation}
\text{RDI}(m) = \max_{r} D^{(r)}(m) - \min_{r} D^{(r)}(m)
\label{eq:rdi}
\end{equation}

A high RDI indicates that the model's perceived value depends heavily on the stakeholder perspective. Table~\ref{tab:rdi} presents the RDI for each model.

\begin{table}[t]
\centering
\caption{Role Divergence Index (CIP layer). Highest and lowest scoring roles shown.}
\label{tab:rdi}
\small
\begin{tabular}{@{}l r l l@{}}
\toprule
\textbf{Model} & \textbf{RDI} & \textbf{Best Role} & \textbf{Worst Role} \\
\midrule
Qwen3-Coder      & 31.1 & AI Actor (77.9) & CISO (45.2) \\
GPT-5.4          & 30.8 & AI Actor (79.2) & CISO (48.4) \\
Qwen3-Plus       & 30.1 & AI Actor (81.2) & CISO (51.1) \\
Opus 4.6         & 29.2 & AI Actor (80.2) & CISO (51.0) \\
Grok Code Fast   & 25.1 & AI Actor (83.8) & CISO (58.7) \\
Gemini 2.5 Flash & 23.6 & AI Actor (84.9) & Researcher (61.1) \\
Sonnet 4.6       & 19.9 & AI Actor (85.6) & Researcher (64.2) \\
Gemini 3.1 Pro   & 21.9 & AI Actor (85.7) & Eng. (63.8) \\
Gemini 2.5 Pro   & 20.1 & AI Actor (86.3) & Researcher (65.2) \\
Kimi K2.5        & 21.3 & AI Actor (86.4) & Eng. (65.1) \\
Haiku 4.5        & 16.8 & AI Actor (85.9) & CAIO (69.1) \\
Gemini 3 Flash   & 21.3 & AI Actor (87.5) & Eng. (66.2) \\
\bottomrule
\end{tabular}
\end{table}

Qwen3-Coder has the highest RDI at 31.1 points: it earns A for Head of Engineering (76.3) and AI Actor (77.9) but D for CISO (45.2). GPT-5.4 shows a similar pattern with 30.8 points of divergence. These models have conservative prediction strategies (high precision, low recall) that satisfy engineering and autonomy lenses but fail the CISO's severity-weighted recall requirements.

Haiku 4.5 has the lowest RDI at 16.8 points, indicating the most balanced profile across stakeholder perspectives. Its strong performance on both detection accuracy and operational dimensions produces consistent grades (B across all non-AI-Actor roles).

\begin{figure}[t]
\centering
\includegraphics[width=\linewidth]{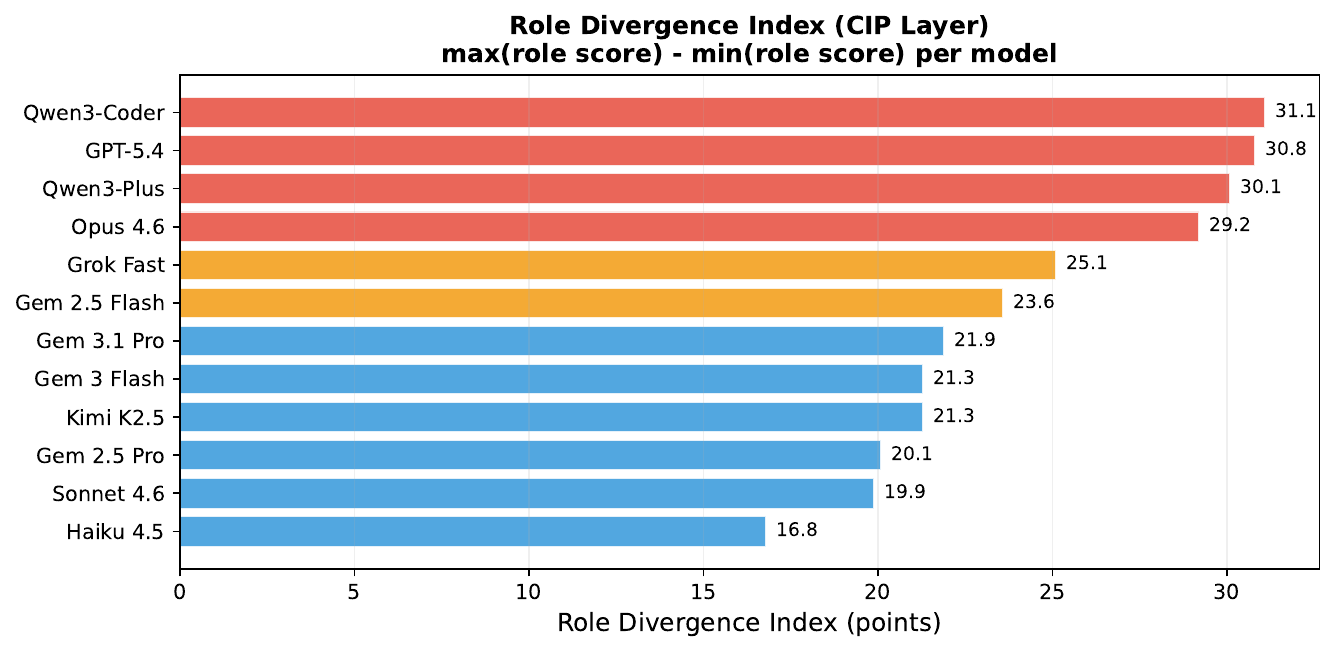}
\caption{Role Divergence Index per model (CIP layer). Red = high divergence ($\geq$28), yellow = moderate ($\geq$22), blue = low. Higher RDI means the model's value depends more on stakeholder perspective.}
\label{fig:rdi}
\end{figure}

The worst-scoring role is nearly always the CISO (for 7 of 12 models), while the best-scoring role is always the AI Actor. This structural asymmetry reflects the current state of frontier models: parse reliability and format compliance (which the AI Actor lens weights heavily) are largely solved, while severity-aware vulnerability detection (which the CISO lens weights heavily) remains challenging. Figure~\ref{fig:rdi} visualizes the RDI distribution.

\subsection{CIP vs.\ Tool-Use Layer Analysis}
\label{sec:cip_vs_tu}

We evaluated models on both CIP and TU layers. Tool-Use runs show several consistent patterns:

\begin{itemize}[noitemsep,topsep=2pt]
    \item \textbf{Lower leaderboard scores.} The top CIP score is 49.6\% (Gemini 3 Flash) vs.\ 45.8\% (Gemini 3.1 Pro) for TU. Tool-use adds complexity: models must navigate repositories, choose which files to read, and synthesize information across multiple turns.
    \item \textbf{Higher cost.} TU runs are 10--100$\times$ more expensive than CIP. Gemini 3.1 Pro costs \$0.111/task in CIP but \$1.578/task in TU. Claude Haiku 4.5 costs \$0.005/task in CIP but \$0.256/task in TU.
    \item \textbf{Additional dimensions available.} TU runs enable location accuracy (D7, D8) and tool-use dimensions (D24--D27), providing a richer signal for roles like Security Researcher and AI-as-Actor.
\end{itemize}

Table~\ref{tab:tu_scores} presents the TU layer role scores for all 11 models evaluated on tool-use. Comparing against the CIP scores in Table~\ref{tab:decision_scores} reveals that TU scores are generally lower for the CISO and Researcher but more stable for AI Actor. Gemini 3.1 Pro Preview achieves the highest TU CISO score (B, 69.4), surpassing its CIP CISO score (B, 67.5), one of the few models where TU improves role-specific performance.

\begin{table*}[t]
\centering
\caption{Per-role Decision Scores (TU layer). Grade thresholds: A $\geq$ 75, B $\geq$ 60, C $\geq$ 50, D $\geq$ 40, F $<$ 40. Sorted by TU leaderboard score.}
\label{tab:tu_scores}
\small
\begin{tabular}{@{}l c c c c c c@{}}
\toprule
\textbf{Model} & \textbf{LB \%} & \textbf{CISO} & \textbf{CAIO} & \textbf{Researcher} & \textbf{Head Eng.} & \textbf{AI Actor} \\
\midrule
Gemini 3.1 Pro Preview   & 45.8 & B (69.4) & C (56.1) & B (65.0) & D (44.5) & A (75.4) \\
Gemini 3 Flash Preview   & 44.2 & B (68.9) & C (55.9) & B (64.1) & D (44.5) & B (73.7) \\
Claude Sonnet 4.6        & 42.1 & B (66.3) & C (56.1) & B (60.6) & D (43.9) & B (74.7) \\
Claude Opus 4.6          & 39.0 & C (56.0) & C (55.6) & C (53.9) & D (43.8) & B (71.7) \\
Claude Haiku 4.5         & 37.8 & B (66.5) & C (57.1) & B (60.3) & D (47.8) & B (72.8) \\
Gemini 2.5 Pro           & 35.2 & C (59.7) & B (64.4) & C (55.2) & C (52.0) & A (79.7) \\
Qwen3-Coder-Plus         & 34.9 & C (56.4) & B (63.7) & C (50.9) & C (54.1) & A (76.7) \\
Grok Code Fast 1         & 34.4 & C (55.5) & B (64.3) & C (50.9) & C (50.6) & A (76.6) \\
Gemini 2.5 Flash         & 34.2 & C (56.0) & B (67.4) & C (52.4) & C (56.3) & A (81.0) \\
Qwen3-Coder              & 32.5 & D (48.9) & C (59.7) & D (48.2) & C (53.2) & B (70.8) \\
GPT-5.4                  & 31.1 & D (45.2) & B (64.2) & D (48.6) & B (60.8) & A (75.6) \\
\bottomrule
\end{tabular}
\end{table*}

The CIP layer provides a cost-effective baseline evaluation; the TU layer adds depth at a substantial cost premium. For cost-sensitive deployments, CIP evaluation alone may be sufficient, while organizations investing in agent-based security tools should evaluate on TU to capture tool-use quality. The TU layer reveals a key difference from CIP: the Head of Engineering scores drop sharply (most models score D), driven by high cost per task (D18) and slow wall times (D21) inherent to multi-turn tool-use evaluation.

\subsection{Cost and Efficiency Analysis}
\label{sec:cost}

Cost was tracked for 8 of 12 models (Anthropic, Google, and OpenAI). Table~\ref{tab:cost_cip} presents the CIP cost data.

\begin{table}[t]
\centering
\caption{CIP layer cost analysis (8 models with cost tracking).}
\label{tab:cost_cip}
\small
\begin{tabular}{@{}l r r r r@{}}
\toprule
\textbf{Model} & \textbf{\$/task} & \textbf{Total} & \textbf{MCC/\$} & \textbf{TPM} \\
\midrule
Haiku 4.5        & \$0.005 & \$2.11   & 0.019 & 9.8 \\
GPT-5.4          & \$0.007 & \$2.97   & 0.042 & 15.1 \\
Gemini 2.5 Flash & \$0.008 & \$3.23   & 0.031 & 4.1 \\
Sonnet 4.6       & \$0.017 & \$6.76   & 0.025 & 4.4 \\
Opus 4.6         & \$0.019 & \$7.56   & 0.009 & 7.3 \\
Gemini 2.5 Pro   & \$0.033 & \$13.45  & 0.011 & 2.2 \\
Gemini 3 Flash   & \$0.039 & \$15.87  & 0.009 & 0.9 \\
Gemini 3.1 Pro   & \$0.111 & \$44.81  & 0.004 & 0.7 \\
\bottomrule
\end{tabular}
\end{table}

\begin{figure}[t]
\centering
\includegraphics[width=\linewidth]{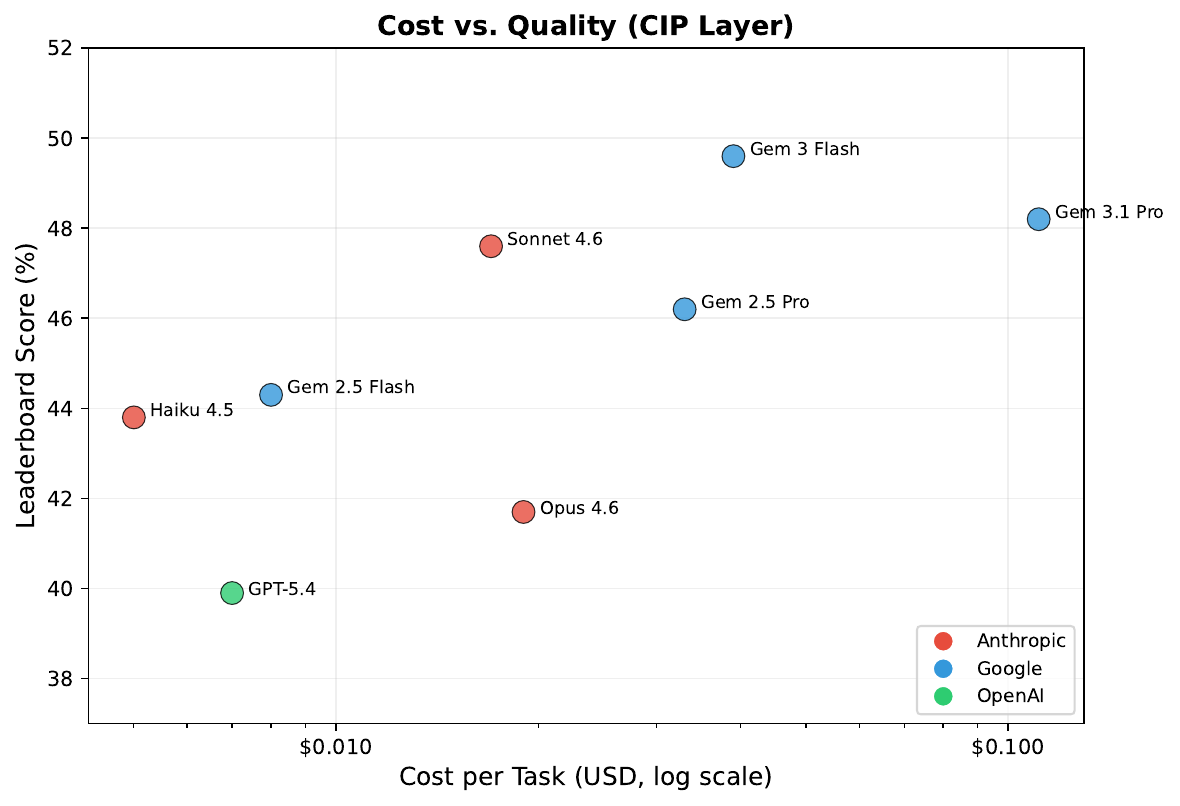}
\caption{Cost per task vs.\ leaderboard score (CIP layer, 8 models with cost tracking). Log-scale x-axis. Higher and further left is better.}
\label{fig:cost_quality}
\end{figure}

Figure~\ref{fig:cost_quality} plots cost per task against leaderboard score for the 8 cost-tracked models, revealing diminishing returns at higher price points.

\textbf{GPT-5.4 delivers the best quality-per-dollar.} At \$0.007/task, GPT-5.4 achieves an MCC/\$ ratio of 0.042, the highest among tracked models. Its low cost makes it attractive for the Head of Engineering lens, which weights D18 (Cost per Task) and D20 (MCC per Dollar).

\textbf{Cost does not predict accuracy.} Gemini 3.1 Pro costs 15$\times$ more than GPT-5.4 per task but scores only 8.3 leaderboard points higher (48.2\% vs.\ 39.9\%). Gemini 3 Flash Preview, the top performer, costs 5.3$\times$ more than GPT-5.4.

\textbf{Throughput varies widely.} GPT-5.4 processes 15.1 tasks per minute while Gemini 3.1 Pro manages only 0.7 TPM, a 21$\times$ gap. For CI/CD integration where latency matters, this difference is significant. The Head of Engineering lens captures this via D22 (Throughput, weight 5).

\textbf{Haiku 4.5 is the cheapest option.} At \$0.005/task (\$2.11 total for 406 tasks), Haiku is the most economical model. Projected to 10K tasks/year, this amounts to roughly \$52. Despite its low cost, Haiku scores B (71.2) for the CISO, outranking models that cost 3--20$\times$ more. This makes Haiku a strong candidate for organizations running frequent security scans on a budget.

\textbf{TU layer costs are substantially higher.} Gemini 2.5 Flash, the cheapest TU option at \$0.011/task, still costs 37\% more than its CIP equivalent. At the high end, Gemini 3.1 Pro costs \$1.578/task in TU (\$639.09 total), a 14$\times$ increase over its CIP cost. Token consumption drives this: TU runs for Haiku 4.5 average 236K tokens/task vs.\ roughly 2.8K in CIP, an 84$\times$ increase due to multi-turn repository navigation.

\textbf{Cost tracking limitation.} The four OpenRouter-routed models (Kimi K2.5, Grok Code Fast 1, Qwen3-Coder, Qwen3-Coder-Plus) did not report cost data. Their cost-derived dimensions (D18, D19, D20) are excluded via the dynamic denominator adjustment in Equation~\ref{eq:decision_score}. This means their CAIO and Head of Engineering scores reflect fewer dimensions than models with cost tracking, which should be considered when comparing across providers.

\subsection{Dimension Sensitivity Analysis}
\label{sec:sensitivity}

To identify which dimensions drive model differentiation for each role, we compute the Impact score:

\begin{equation}
\text{Impact}_i^{(r)} = w_i^{(r)} \cdot \text{Var}_{m}[s_i^{(m)}]
\label{eq:impact}
\end{equation}

This product of weight and cross-model variance identifies dimensions that both matter to the role (high weight) and differentiate between models (high variance).

For the \textbf{CISO}, the highest-impact dimensions are D28 (Severity-Weighted Recall, weight 10) and D29 (Critical Miss Rate, weight 8), which have high cross-model variance because some models detect most critical vulnerabilities while others miss them entirely. D1 (MCC, weight 10) also contributes high impact.

For the \textbf{Head of Engineering}, D3 (Precision, weight 12) and D8 (Actionable Finding Rate, weight 10) drive differentiation. Models with conservative prediction strategies (Qwen3-Coder, GPT-5.4) achieve near-perfect precision, while models with higher recall but lower precision (Gemini 3 Flash) score lower on these dimensions but higher on D2 (Recall, weight 5).

For the \textbf{AI Actor}, variance is low across all Robustness dimensions (most models parse well), so the Detection dimensions D1 (MCC, weight 10) and D4 (F1, weight 7) drive the modest differentiation that exists within the A-grade band.

% ============================================================
% 8. DISCUSSION
% ============================================================
\section{Discussion}
\label{sec:discussion}

\subsection{Limitations}

\textbf{Weight subjectivity.} The weight vectors, while informed by domain expertise and industry frameworks, involve judgment. Different organizations may prioritize differently within the same role. We address this by storing profiles in YAML files that organizations can customize. Adding or modifying a role requires only a new YAML file specifying the dimension subset and weights.

\textbf{Single-run evaluation.} Without multi-run data, we cannot estimate confidence intervals on dimension scores. Dimensions computed from small subsets (e.g., Memory Safety with only 20 tasks, or rare CWE categories) may have high variance not captured in a single run.

\textbf{No SAST FP tasks.} The current dataset contains no SAST false-positive tasks, so D13 (SAST FP Filtering) is always excluded. This dimension would be valuable for evaluating a model's ability to distinguish tool-reported false alarms from real vulnerabilities.

\textbf{Dataset size for rare categories.} While 406 tasks is substantial, some categories have few tasks (Memory Safety: 20, Data Integrity: 36). Per-category metrics for these smaller groups carry wider implicit confidence intervals.

\textbf{Paired task design.} The balanced TP/post-patch design means that True Negative Rate (D5) and related specificity measures may be inflated relative to real-world scanning, where the ratio of vulnerable to non-vulnerable code is far more skewed.

\textbf{Cost tracking gaps.} Cost data was unavailable for 4 of 12 models (those routed via OpenRouter). This limits cost-efficiency analysis to 8 models and means that cost-derived dimensions are excluded for the remaining 4, reducing the effective dimensionality of their role scores.

\textbf{Severity data on TP tasks only.} Severity annotations (Critical, High, Medium, Low) exist only for true positive tasks. Post-patch tasks carry no severity label, so severity-weighted dimensions (D28--D30) are computed over the TP subset only.

\subsection{Implications for Model Selection}

Our results carry practical implications for organizations evaluating LLMs for security use:

\textbf{No universal ``best model'' exists.} Gemini 3 Flash Preview leads the aggregate leaderboard (49.6\%), but Claude Haiku 4.5 is a better choice for a CISO (71.2 vs.\ 73.3) despite ranking 6 places lower overall. GPT-5.4, ranked 11th on the leaderboard, is the top choice for engineering teams (A, 76.7) and delivers the best cost-quality ratio (\$0.007/task, MCC/\$ = 0.042). Model selection must be anchored to the decision-maker's priorities.

\textbf{Conservative models suit engineering; aggressive models suit security.} Models with high precision and low recall (Qwen3-Coder, GPT-5.4) excel for the Head of Engineering, who values actionable findings and low false-positive rates. Models with high recall and broader detection (Gemini 3 Flash, Haiku 4.5) excel for the CISO, who values coverage and severity-aware detection. The same behavioral trait, conservative prediction, is a strength for one stakeholder and a weakness for another.

\textbf{CIP evaluation is sufficient for most decisions.} The CIP layer captures 65--70\% of the framework's discriminative power at 10--100$\times$ lower cost than TU. For organizations primarily concerned with detection quality, reasoning, and cost, CIP-based role scores provide actionable guidance. The TU layer adds value mainly for AI-as-Actor and Security Researcher evaluations, where tool-use and location dimensions matter.

\textbf{Category-specific weaknesses drive CISO failures.} The CISO lens penalizes blind spots. Opus 4.6 scores 0.000 F1 on Authentication Failures; Qwen3-Coder achieves F1 below 0.13 on 4 of 8 categories. These category gaps, invisible in aggregate scores, determine whether a model can be trusted in a security program.

\subsection{Comparison with Concurrent Work}

Several benchmarks published concurrently address related but distinct problems. SecVulEval~\citep{lu2025secvuleval} is the largest single-language benchmark (25K C/C++ samples), but its scope is limited to C/C++ and it produces a single F1 score. SEC-bench~\citep{lee2025secbench} evaluates LLM agents on exploit generation and patching, tasks complementary to the detection focus of \framework. TOSSS~\citep{damie2026tosss} frames detection as binary snippet selection, whereas \seclens requires CWE classification, location, and evidence generation, enabling richer dimension computation.

Our contribution is orthogonal: we do not propose a new detection benchmark, but a scoring layer that transforms any detection benchmark's per-task results into stakeholder-specific evaluations. The \framework dimensions could be computed over SecVulEval or TOSSS results with minimal adaptation, provided those benchmarks emit per-task verdict, CWE, and location fields.

\subsection{Future Work}

\textbf{Custom organizational profiles.} The YAML-based architecture makes it straightforward to define custom roles. An interactive profile editor could help security teams create weight vectors aligned with their specific risk appetite and operational constraints.

\textbf{SAST FP task collection.} Curating a set of SAST-flagged but manually-confirmed non-vulnerable code samples would enable D13 and provide a direct measure of a model's ability to filter static analysis noise.

\textbf{Multi-run evaluation.} Evaluating each model multiple times with different seeds would yield confidence intervals on dimension scores and enable a cross-run stability dimension.

\textbf{Expanded model coverage.} Testing additional open-weight models (e.g., DeepSeek, Llama) and specialized security models would broaden the analysis. Running open-weight models locally would also resolve the cost tracking gap for OpenRouter-routed evaluations.

\textbf{Temporal tracking.} Evaluating successive model versions (e.g., Claude Sonnet 4.5 $\to$ 4.6) on the same dataset would enable regression risk assessment and capability progress tracking across role lenses.

\textbf{Cross-benchmark integration.} Applying the \framework scoring layer to results from other benchmarks (SecVulEval~\citep{lu2025secvuleval}, SEC-bench~\citep{lee2025secbench}, TOSSS~\citep{damie2026tosss}) would test whether role-specific divergence generalizes beyond the \seclens dataset and task format.

% ============================================================
% 9. CONCLUSION
% ============================================================
\section{Conclusion}
\label{sec:conclusion}

We have presented \framework, a multi-stakeholder evaluation framework for LLM-based security vulnerability detection. The framework defines 35 shared evaluation dimensions across 7 categories, applied through 5 role-specific weight profiles where each role selects 12--16 dimensions with weights summing to 80.

We evaluated 12 frontier models on 406 tasks spanning 93 projects, 10 programming languages, and 8 OWASP-aligned vulnerability categories. The results demonstrate that single-score benchmarks obscure information that matters for organizational decisions:

\begin{itemize}[noitemsep,topsep=2pt]
    \item Decision Scores diverge by up to 31 points across roles for the same model. Qwen3-Coder earns A (76.3) for Head of Engineering but D (45.2) for CISO.
    \item GPT-5.4 earns A for Head of Engineering (76.7) but D for CISO (48.4), driven by its conservative prediction strategy (high precision, low recall on critical vulnerabilities).
    \item Leaderboard rank does not predict role-specific rank. Claude Haiku 4.5, ranked 8th overall, scores 2nd for CISO (71.2).
    \item No single model dominates all vulnerability categories. Six different models lead at least one of the 8 OWASP-aligned categories.
    \item The AI Actor lens shows all models earning A grades, while the CISO lens shows grades ranging from D to B, confirming that model robustness is largely solved but severity-aware detection remains challenging.
\end{itemize}

The YAML-based weight profiles enable organizational customization: teams can adjust or create roles to match their specific risk appetites and operational priorities. The framework integrates directly with the \seclens evaluation pipeline, consuming existing per-task result records without requiring any modifications to the underlying benchmark.

We release the full implementation as open-source software to support organizational adoption and community extension.

% ============================================================
% ACKNOWLEDGMENTS
% ============================================================
\section*{Acknowledgments}

The authors thank the open-source LLM and security research communities for the foundational work that enables this research.

% ============================================================
% REFERENCES
% ============================================================

\end{document}